\documentclass[journal abbreviation]{copernicus2}

\begin{document}

\title{Anisotropic J\"uttner (relativistic Boltzmann) distribution
}

\author[1]{R. A. Treumann\thanks{Visiting the International Space Science Institute, Bern, Switzerland}}
\author[2]{W. Baumjohann}

\affil[1]{Department of Geophysics and Environmental Sciences, Munich University, Munich, Germany}
\affil[2]{Space Research Institute, Austrian Academy of Sciences, Graz, Austria}

\runningtitle{J\"uttner-Function}

\runningauthor{R. A. Treumann and W. Baumjohann 
}

\correspondence{R. A.Treumann\\ (rudolf.treumann@geophysik.uni-muenchen.de)}

\received{ }
\revised{ }
\accepted{ }
\published{ }


\firstpage{1}

\maketitle

{\subsection*{\bf{Abstract}} 
The J\"uttner (covariant Boltzmann) distribution is provided for anisotropic pressure (or temperature) tensors. Its manifestly covariant form follows straightforwardly from its scalar property.}
\vspace{0.5cm}

The J\"uttner distribution \citep{juttner1911a} is the relativistically generalised isotropic Maxwell-Boltzmann distribution, whether written in its dependence on relativistic particle energy $\epsilon_\mathbf{p}=m\gamma(\mathbf{p})c^2$ or momentum $\mathbf{p}$, with $\gamma=\sqrt{1+\mathbf{p}^2/m^2c^2}$. The phase-volume element $\mathrm{d}\mathbf{x}\:\mathrm{d}\mathbf{p}$ is covariant (a consequence of its scalar nature). Hence the J\"uttner distribution is as well covariant but not manifestly covariant. In anisotropic relativistic gases or plasmas the form of the J\"uttner distribution is usually assumed \citep[for recent examples cf., e.g.,][]{swisdak2013,lopez2014,alves2015,zenitani2015,vore2015}. Below we provide its simple analytical derivation and manifestly covariant version. 

Maxwell-Boltzmann distributions are solutions of the stationary one-particle Boltzmann equation with argument the ratio of the single particle energy to average thermal energy, viz. $\epsilon_\mathbf{p}/T$, with $T$ temperature in energy units. Properly normalised they give the probability at temperature $T$ for finding all particles of given momentum $\mathbf{p}$ (or energy $\epsilon_\mathbf{p}$) in the interval d$\mathbf{p}$ (or d$\epsilon_\mathbf{p}$) in the momentum-space volume d$\mathbf{p}$. 
With momentum vector $\mathbf{p}=(p_\perp\cos\phi,\: p_\perp\sin\phi,\: \,p_\|)$ in index notation
\begin{equation}\label{eq-three}
\epsilon_\mathbf{p}^2 = c^2p_i\delta^i_j p^j +m^2c^4
\end{equation}
suggests introduction of a temperature anisotropy guided by the diagonal anisotropy of pressure $\textsf{P}=  N\big[T_\perp\delta_i^j +(T_\|-T_\perp)\delta_3^3\big]$ (as for instance in magnetised plasma), with anisotropy in direction $3$ (in plasma the direction of the magnetic field $\mathbf{b}=\mathbf{B}/B$, for instance). The inverse pressure/temperature tensor is $\textsf{P}^{-1} =\left(T_\perp N\right)^{-1}\mathbf{\Theta}$,
\begin{equation}
\mathbf{\Theta}=\Theta^i_j=\delta_i^j+(A-1)\delta_3^3\quad \mathrm{with}\quad A=T_\perp/T_\|
\end{equation}
Replace $\delta^i_j$  in (\ref{eq-three}) with  $\Theta^i_j$,  valid in the 4-velocity frame $U_j=(\epsilon_\mathbf{p}/mc,0)$, 
and put $p_\perp=p\:\sin\theta,p_\|=p\:\cos\theta$ yields
\begin{equation}\label{eq-ept}
\frac{\epsilon^2_\mathbf{p}}{T_\perp^2}=\frac{m^2c^4}{T_\perp^2}{\left[1+\frac{p^2}{m^2c^2}\Big(\sin^2\theta+A\cos^2\theta\Big) \right]}
 \end{equation}

The square root of (\ref{eq-ept}) enters the Boltzmann factor. Define $\beta_\perp=mc^2/T_\perp$. Up to normalisation $C$, the anisotropic J\"uttner distribution function of the ideal gas becomes
\begin{equation} \label{jutt}
F(\mathbf{p})=C \exp\Bigg\{-\beta_\perp\sqrt{1+\frac{p_\perp^2}{m^2c^2}+A\frac{p_\|^2}{m^2c^2}}~\Bigg\}
\end{equation}
With $A=1, \beta_\perp=\beta_\|=\beta$ this is the ordinary J\"uttner function. Expanding the root in the limit ${c\to\infty}$ reproduces the ordinary nonrelativistic anisotropic Maxwell-Boltzmann distribution.  Extensions to non-ideal gases are straightforward. 

The above, in principle trivial, result differs from earlier ones.  
Normalization, the purpose of J\"uttner's effort, yields   
\begin{equation}
C= N\sqrt{A}\beta_\perp/4\pi(mc)^3K_2(\beta_\perp)
\end{equation}
with $K_2(\beta_\perp)$ the Bessel function. This trivially contains the anisotropy factor $A$. The above covariant result is valid in time-like slices of real configuration space.  Manifestly covariant isotropic versions have been provided as well \citep[cf.,][]{acosta2010,curado2016}. Since $F(\mathbf{p})$ is a scalar phase space density, its manifestly covariant version is $F(x^\nu,p^\nu)\sqrt{-g}$ for both isotropic and anisotropic cases. $g<0$ is the determinant of the metric tensor $g_{\mu\nu}$ in $(+---)$ metric, a version to be applied in curvilinear coordinates. In general relativistic 4-space, $\mu,\nu=0,1,2,3$, and $\mathbf{p}\to p^\nu$ is the 4-momentum.

J\"uttner's anisotropic distribution is useful for analytical calculations. In numerical (PIC) simulations the initial distribution is prescribed. In practice there is little need to choose it to satisfy the J\"uttner equilibrium requirement. Solving for all relativistic particle orbits in selfconsistent fields it adjusts itself to the physical distribution that evolves under the mutual interactions. 

This derivation indicates that in relativistic media the isotropic temperature $T$ and its inverse $\beta=1/T$ should be understood as vectors \citep[as suggested for different reasons by][]{nakamura2009}. For anisotropy they become tensors. 


\end{document}